\documentclass[12pt, a4paper]{article}
\usepackage{a4wide,amssymb,amsmath,amscd,graphicx}
\usepackage[all]{xy}

\newcommand{\Z}{{\mathbb Z}}
\newcommand{\R}{{\mathbb R}}
\newcommand{\C}{{\mathbb C}}

\renewcommand{\Re}{\mathrm{Re}}
\renewcommand{\Im}{\mathrm{Im}}

\begin{document}

\topmargin -2pt

\headheight 0pt

\topskip 0mm \addtolength{\baselineskip}{0.20\baselineskip}
\begin{flushright}
{\tt KIAS-P09044}
\end{flushright}


\begin{center}
{\Large \bf Morita Equivalence of Noncommutative Supertori } \\

\vspace{10mm}

{\sc Ee Chang-Young}\footnote{cylee@sejong.ac.kr}\\
{\it Department of Physics, Sejong University, Seoul 143-747, Korea}\\

\vspace{5mm}

{\sc Hoil Kim}\footnote{hikim@knu.ac.kr}\\

{\it Department of Mathematics, Kyungpook National University,\\
Taegu 702-701, Korea}\\

\vspace{2mm}

and \\

\vspace{2mm}

{\sc Hiroaki Nakajima}\footnote{nakajima@skku.edu}\\

{\it Department of Physics and Institute of Basic Science \\
Sungkyunkwan University, Suwon 440-746, Korea}\\

\vspace{10mm}

{\bf ABSTRACT} \\
\end{center}

\vspace{2mm}

\noindent
 In this paper we study the extension of Morita equivalence
 of noncommutative tori to the supersymmetric case.
The structure of the symmetry group yielding Morita equivalence
 appears to be intact but its parameter field becomes supersymmetrized
 having both body and soul parts.
 Our result is mainly in the two dimensional case in which
 noncommutative supertori have been constructed recently:
 The group $SO(2,2,{\cal V}_{\Z}^0)$, where ${\cal V}_{\Z}^0$ denotes Grassmann even number
 whose body part belongs to ${\Z}$,
 yields Morita equivalent noncommutative supertori in two dimensions.
  \\

\setcounter{footnote}{0}



\thispagestyle{empty}

\newpage
\section{Introduction}

In the work of Connes, Douglas, and Schwarz \cite{cds}
Morita equivalence of two dimensional noncommutative tori
resulting from toroidal compactifications of M(atrix) theory
 was mentioned in relation with
T-duality in string theory \cite{jp,gsw}.
%
Then it was proved by Rieffel and Schwarz \cite{rfsz98}
that the actions of the group $SO(n,n,{\Z})$ on an antisymmetric $n\times n$ matrix
$\Theta$ which represents noncommutativity parameters for an $n$-dimensional noncommutative
torus yield Morita equivalent tori.
Then Schwarz \cite{schwarz98} showed that compactifications on Morita equivalent
tori are physically equivalent, corresponding to T-duality in string theory.

Recently, Berkovits and Maldacena \cite{bm08} showed that tree level superstring theories
on certain supersymmetric back grounds are related by the shift symmetry of a certain fermionic
coordinate, which they called ``fermionic" T-duality.
This is very similar to  bosonic T-duality in the sense that it
is a symmetry under a shift of a coordinate.
Bosonic T-duality is related with torus compactification.
This naturally leads us to an expectation that the
fermionic T-duality might be related to supertorus compactification.
On the other hand, the fermionic T-duality relates different RR field backgrounds
\cite{bm08}.
 Furthermore, the background RR field leads to
a non(anti)commutative superspace \cite{Se}.
This makes us think that the fermionic T-duality might be related with
the compactification on  noncommutative supertorus rather than on commutative supertorus.
In order to further investigate this idea, we have to understand
 Morita equivalence of noncommutative supertori first,
since in the bosonic case it has been shown that
T-duality corresponds to Morita equivalence
of noncommutative tori \cite{schwarz98}.
As a step in that direction, we will study the Morita equivalence of noncommutative supertori
in this paper.
%

Commutative supertorus was constructed by Rabin and Freund \cite{rf1988} based on the
work of Crane and Rabin \cite{cr1988} on super Riemann surfaces.
The supertorus was obtained as a quotient of superplane by a subgroup of superconformal
group Osp$(1|2)$
which acts properly discontinuously on the plane together with the metrizable condition.
These two conditions boil down to proper latticing of the superplane,
and can be expressed as appropriate translation properties along the cycles of the torus.
Noncommutative tori can be constructed by embedding a lattice \cite{rief88,manin1-tr,manin3,ek08}
into Heisenberg group \cite{mumford,sthan98,jrosenberg}.
The lattice embedding determines how the generators of noncommutative torus,
which correspond to the translation operators along the cycles of commutative torus,
would act on the module of noncommutative torus.

Based on the construction of super Heisenberg group \cite{ekn07}
as a central extension of ordinary superspace,
the embedding maps for noncommutative supertori in two dimensions were obtained in \cite{ekn08}.
Based on this construction of noncommutative supertori, here we study Morita equivalence
of noncommutative supertori
by investigating symmetry actions that give dual embedding maps, i.e., finding
the endomorphisms of the module that yield
Morita equivalent tori.

The paper is organized as follows.
In section 2,
we review Morita equivalence in relation with the group $SO(n,n,{\Z})$
in the bosonic noncommutative $n$-tori case.
In section 3, we briefly
 recall the construction of noncommutative supertori,
then consider the Morita equivalence of noncommutative supertori
in a general setting and identify symmetry operators yielding Morita equivalent tori.
When restricted to the two dimensional case, this yields the symmetry group $SO(2,2,{\cal V}_{\Z}^0)$
 where ${\cal V}_{\Z}^0$ denotes Grassmann even number
 whose body part belongs to ${\Z}$.
We conclude in section 4.
\\

%
\section{Morita equivalence in the bosonic case } \label{Morita-boson}

\noindent
In this section we review Morita equivalence of noncommutative tori
in the bosonic case in line with the work of Rieffel and Schwarz \cite{rfsz98}.
In general, Morita equivalence relates two algebras ${\cal A}$ and $\hat{\cal A}$ in such a way that
for every ${\cal A}$-module $R$ there exists $\hat{\cal A}$-module $\hat{R}$
where the correspondence $R \rightarrow \hat{R}$ is an equivalence of categories of
${\cal A}$-modules and $\hat{\cal A}$-modules.
More specifically, (strong) Morita equivalence of $C^*$-algebra can be defined as follows.
If we consider a finite projective right module $\cal{E}$ over a $C^*$-algebra $A$,
then the algebra $End_A {\cal E}$ of endomorphisms of  ${\cal E}$ has a canonical
structure as a $C^*$-algebra.
We say that a $C^*$-algebra $A'$ is (strongly) Morita equivalent to $A$ if it is isomorphic to
 $End_A {\cal E}$  for some finite projective module ${\cal E}$.

An $n$-dimensional noncommutative torus($A_\theta^n$) is an associative algebra with involution having
unitary generators $U_1, \dots , U_n$ obeying the relations
\begin{align}
U_iU_j=e^{2 \pi i \theta_{ij}}U_jU_i , ~~~ i,j=1,\dots,n,
\label{nctorus}
\end{align}
where $(\theta_{ij})$ form a real $n\times n$ anti-symmetric matrix $\Theta$.

Let ${\cal D}$ be a lattice in ${\cal G}=M\times \hat{M}$, where
$M={\R}^p\times{\Z}^q$ with $2p+q=n$ and $\hat{M}$ is its dual.
The embedding map $\Phi$ under which ${\cal D}$ is the image of ${\Z}^n$
 determines a projective module $E$ on which the algebra of noncommutative torus acts.
In the Heisenberg representation the operators $U$'s are
 defined by
\begin{equation}
U_{(m,\hat s)}f(r)=e^{2\pi i <r, \hat s>}f(r+m), ~~ m,r \in M, ~ \hat s \in \hat{M}, ~
  f \in E ,
\label{heisrep}
\end{equation}
where $<r, \hat s>$ is a usual inner product between $M$ and $\hat{M}$.
%
%
In this representation, we can easily get the antisymmetric matrix $\Theta$ characterizing noncommutative torus
as follows.
\begin{eqnarray}
U_{(m,\hat s)}U_{(n,\hat t)}f(r) & = & e^{2\pi i (<r, \hat s+\hat t>+<m, \hat t>)}f(r+m+n), \nonumber \\
 & = & e^{2\pi i <m, \hat t>} U_{(m+n,\hat s + \hat t))}f(r),
\label{cocycleb}
\end{eqnarray}
 where $ m,n,r  \in M, ~ \hat s, \hat t  \in \hat{M}, ~  f \in E $, and thus
\begin{eqnarray}
U_{(m,\hat s)}U_{(n,\hat t)} =   e^{2\pi i (<m, \hat t> - <n, \hat s>)}U_{(n,\hat t)}U_{(m,\hat s)} .
\label{cocycler}
\end{eqnarray}
The Heisenberg cocycle $\beta$ is defined by
 \begin{equation}
\label{heis-cocycle}
\beta ((m,\hat{s}),(n,\hat{t})) = \exp ( 2 \pi i <m, \hat{t}>) ,
\end{equation}
and the corresponding skew cocycle $\varphi$ is defined by
\begin{equation}
\label{rho-cocycle}
\varphi ((m,\hat{s}),(n,\hat{t})) = \exp ( 2 \pi  i (<m,\hat{t}> - <n,\hat{s}> )).
\end{equation}
 Denoting $\vec{e}_i :=(m,\hat s), \; \vec{e}_j :=(n,\hat t)$,  $\theta_{ij} $ can be expressed as follows.
 \begin{align}
\theta_{ij} = \vec{e}_i \cdot  J_0  \vec{e}_j , \ \ {\rm where } \  \ J_0 =\begin{pmatrix} 0 & I \\ -I & 0 \end{pmatrix} .
\label{cyctheta}
\end{align}
With the embedding map $ \Phi =( \vec{e}_1, \vec{e}_2, \cdots, \vec{e}_n )$,
 $\Theta$ is given by
 \begin{align}
\Theta = \Phi^t  J_0  \Phi, \ \ {\rm where } \  \ J_0 =\begin{pmatrix} 0 & I \\ -I & 0 \end{pmatrix} .
\label{Theta}
\end{align}

In order to consider the endomorphisms of $A_\theta^n$, we first consider
the group $O(n,n,{\R})$ on the space of ${\cal T}_n$ of real antisymmetric $n \times n$ matrices.
The group $O(n,n,{\R})$ can be considered as a group of linear transformations of the space ${\R}^{2n}$
preserving the quadratic form $x_1 x_{n+1}+x_2 x_{n+2}+ \cdots + x_n x_{2n}$.
It is convenient to consider coordinates in ${\R}^{2n}$ as two $n$-dimensional vectors
$(a^1, \dots, a^n, b_1, \dots, b_n)$ so that the quadratic form on ${\R}^{2n}$ can be
written as $a^i b_i$. Thus we write the elements of $O(n,n,{\R})$ in $2 \times 2$ block form
%
\begin{equation}
\label{gpelement}
 g =\begin{pmatrix} A& B\cr  C&D
  \end{pmatrix} .
\end{equation}
The blocks $A,B,C,D$ are $n \times n$ matrices satisfying
 $A^t C + C^t A  = B^t D + D^t B=0, \, A^t D + C^t B =1,$ where ${}^t$ denotes
 transpose. The action of $O(n,n,{\R})$ on the space ${\cal T}_n$  is defined by the
 formula
\begin{equation}
 \Theta' = g \Theta:= (A \Theta + B)(C \Theta +D)^{-1} .
 \label{gaction_theta}
\end{equation}

Now, let
\begin{equation}
{\cal T}_n^0 =\{ \Theta \in {\cal T}_n : \; g \Theta {\rm \; \;  is \; defined \; for \; all \; } g \in SO(n,n,{\Z}) \}.
\end{equation}
Then the following theorem holds \cite{rfsz98}.\\
%
%
%
%
{\bf Theorem.} \ {\it For $\Theta \in {\cal T}_n^0$ and $g \in SO(n,n,{\Z})$ the noncommutative torus corresponding
to $g\Theta$ is Morita equivalent to the noncommutative torus corresponding to $\Theta$.} \\
%

 For the proof of the above theorem, we first consider a suitable set of generators for $SO(n,n,{\Z})$.
In \cite{rfsz98}, it was shown as a lemma that
  the three elements, $\rho (R), \nu (N),$ and $\sigma_2$, which we describe below
 generate the group  $SO(n,n,{\Z})$. \\
1) For every matrix $R \in GL(n,{\Z})$, $\rho (R) \in SO(n,n,{\Z})$ defines the following transformations
 \begin{equation}
 \label{rot}
 a'^i=R_j^i a^j, \; \; b'_i = (R^{-1})_i^j b_j .
 \end{equation}
The action of  $\rho (R)$ on $x =(a,b)$ with $a, b  \in  {\R}^{n}$ can be simply expressed  as
 \begin{equation}
 \label{rotm}
 \rho (R)x = \begin{pmatrix} R^t & 0 \\ 0 & R^{-1} \end{pmatrix} \begin{pmatrix} a \\ b \end{pmatrix}.
 \end{equation}
 2) For an antisymmetric $n \times n$ matrix $N$ whose entries are $n_{ij} \in {\Z}$ the transformation $\nu (N) \in SO(n,n,{\Z})$
  is defined as follows.
  \begin{equation}
 \label{shift}
 a'^i=a^i + n^{ij}b_j, \; \; b'_i = b_i .
 \end{equation}
3) For every integer $k$ an element $\sigma_k \in O(n,n,{\Z})$ defines the following transformations
\begin{align}
a'^i = b_i & \; \; {\rm for} \; 1 \leq i \leq k, \; \; a'^i = a^i  \; \; {\rm for} \; \, k < i \leq n , \notag \\
b'_i = a^i & \; \; {\rm for} \; 1 \leq i \leq k, \; \; b'_i = b_i  \;  \; {\rm for} \; \, k < i \leq n  .
\label{inverse}
\end{align}
We refer the proof of the above lemma to Ref. \cite{rfsz98}, and proceed to the proof of the theorem.
The proof of the theorem was done by showing that each action of the above generators of $SO(n,n,{\Z})$
yields Morita equivalent torus.

 For $g=\rho (R)$, the noncommutative torus determined by $\Theta' =R \Theta R^t$ is Morita equivalent
 to the torus corresponding to $\Theta $. This is because the two tori are isomorphic as we see below.
 For a general embedding, one can express an embedding vector $\vec{x}$ in terms of basis vectors $\{\vec{e}_i \}$ as
 $\vec{x}= \sum_1^n x_i \vec{e}_i$ where
 $x_i \in {\Z} $, and $\vec{e}_i$'s satisfy the relation \eqref{cyctheta}.
 Then $\theta_{xy}$ for two embedding vectors $\vec{x},\vec{y}$ is given by
  \begin{equation}
 \label{gentheta}
\theta_{xy} = \sum_{i,j=1}^n ( x_i \vec{e}_i) \cdot   J_0  \ ( y_j \vec{e}_j )  = \sum_{i,j=1}^n x_i \theta_{ij} y_j .
 \end{equation}
Namely,  we get
 \begin{equation}
 \label{gcytheta}
U_x U_y = \exp( 2 \pi i \, x^t  \Theta y) U_y U_x ,
 \end{equation}
where $x=(x_1, \dots, x_n)$ and $ y=(y_1,\dots, y_n)$ belong to $ {\Z}^n $.
Under the action of $\rho (R)$, the transformed $x$ is given by $ x'=  \rho(R) x = R^t x$,
and thus
  \begin{equation}
 \label{trstheta}
U_{x'} U_{y'}  = \exp( 2 \pi i \, (R^t x)^t \Theta (R^t y)) U_{y'} U_{x'}
  =  \exp( 2 \pi i \, x^t R \Theta R^t y)U_{y'} U_{x'} .
 \end{equation}
Therefore, the two tori with $\Theta$ and $\Theta' = R \Theta R^t $ are isomorphic.

 For $\nu (N) \in  SO(n,n,{\Z})$, $\Theta' = \nu (N) \Theta $ is given by replacing $\theta_{ij}$ with
 $\theta'_{ij} = \theta_{ij} + n_{ij}$ for $n_{ij} \in {\Z}$. This does not change the commutation
 relations among $U_x$, therefore $A_{\theta'}^n$ and $A_\theta^n$ correspond to the same
 noncommutative torus.

To prove that $A_{\theta'}^n$ with $\Theta' = \sigma_2 \, \Theta $ is Morita equivalent to $A_\theta^n$,
we first prove it for any  $\sigma_{2p}$ with a positive integer $p$ obeying $2p \leq n $.
 For this we consider $ \Theta \in {\cal T}_n$ in $2 \times 2$ block form whose top left part is a $2p \times 2p$ matrix
denoted by $\theta_{11}$,
\begin{equation}
\label{theta-sigma2p}
\Theta := \begin{pmatrix} \theta_{11} & \theta_{12} \cr \theta_{21} & \theta_{22} \end{pmatrix} .
\end{equation}
Then
\begin{equation}
\label{sigma2p}
\Theta' =\sigma_{2p} \, \Theta
= \begin{pmatrix} \theta_{11}^{-1} & - \theta_{11}^{-1} \theta_{12} \cr
  \theta_{21}\theta_{11}^{-1} & \theta_{22} - \theta_{21} \theta_{11}^{-1}\theta_{12} \end{pmatrix},
\end{equation}
and the noncommutative tori corresponding to $\Theta'$ and $\Theta$ are Morita equivalent.\\
To prove this we first choose an invertible matrix $T_{11}$ such that
$T_{11}^t J_0 T_{11} = - \theta_{11}$ where
\[
J_0 = \begin{pmatrix} 0 &  I_p  \cr - I_p & 0 \end{pmatrix} .
\]
Then set $T_{31} = \theta_{12}^t,$ and $T_{32}$ be any $q \times q$ matrix such that
$\theta_{22} = T_{32}^t - T_{32}$ where $q := n - 2p$.
Now, set a $(n+q) \times n$ matrix $T$ as
 \begin{equation}
\label{T-matrix}
 T := \begin{pmatrix} T_{11} & 0 \cr 0 & I_q \cr T_{31} & T_{32} \end{pmatrix} ,
\end{equation}
and a $(n+q) \times (n+q)$ square matrix  $J$ as
 \begin{equation}
\label{J-matrix}
 J := \begin{pmatrix} J_0 & 0 & 0 \cr 0 & 0 & I_q \cr 0 & - I_q & 0 \end{pmatrix} ,
\end{equation}
then it can be shown that $ T^t J T = - \Theta $.

In \cite{rief88}, the endomorphism algebra is given by  ${\cal S}({\cal D}^\bot , \beta)$
where
\begin{equation}
\label{D-pub}
{\cal D}^\bot = \{ w \in {\cal G} : \varphi (w,z) =1  \; \; {\rm for \; \;  all} \; \; z \in {\cal D} \}  ,
\end{equation}
and $\beta$ is restricted to ${\cal D}^\bot$.
 For a natural isomorphism from ${\Z}^n$ to this ${\cal D}^\bot$, we define an invertible $(n+q) \times (n+q)$ matrix
\begin{equation}
\label{Tbar-matrix}
 \bar{T} := \begin{pmatrix} T_{11} & 0 & 0 \cr 0 & I_q & 0 \cr T_{31} & T_{32} &  I_q  \end{pmatrix} .
\end{equation}
It can be checked that $T^t J x \in {\Z}^n $ exactly if $\bar{T}^t J x \in {\Z}^{n+q}$.
Since $\bar{T}, J$ are invertible, the following holds viewed in $\cal{G}$.
\begin{equation}
\label{dpub-form}
 {\cal D}^\bot = ( \bar{T}^t J)^{-1}( {\Z}^{n+q}) .
\end{equation}
Since the inverse of  $( \bar{T}^t J)$ is given by
 \begin{equation}
\label{Tbar-mult}
 ( \bar{T}^t J)^{-1} = \begin{pmatrix} - J_0 (T_{11}^t)^{-1} & 0 & J_0 (T_{11}^t)^{-1} T_{31}^t  \cr
    0 & 0 & -I_q \cr  0 & I_q & - T_{32}^t \end{pmatrix} ,
\end{equation}
 we get  the following desired embedding map
\begin{equation}
\label{embedmap tpub}
 S = \begin{pmatrix}  J_0 (T_{11}^t)^{-1} & - J_0 (T_{11}^t)^{-1} T_{31}^t  \cr
    0  & I_q \cr  0 &   T_{32}^t \end{pmatrix} ,
\end{equation}
which gives an isomorphism from ${\Z}^n$ onto ${\cal D}^\bot$.
One can now easily show that
\begin{equation}
\label{trans theta}
  S^t J  S  = \sigma_{2p} \, \Theta .
\end{equation}
Note that here we have a freedom to choose the size of the component $\theta_{11}$ in \eqref{theta-sigma2p},
which is given by a $2p \times 2p$ matrix, for any $p$ within the range of  $ 2 \leq 2p \leq n $.
Thus, the noncommutative tori
corresponding to $\sigma_{2p} \, \Theta$
and $\sigma_{2} \, \Theta$ are Morita equivalent, and this proves the last part of the  theorem.

\section{Noncommutative supertori and Morita equivalence} \label{Morita-super}

In this section, we first briefly review the construction of noncommutative
supertori in two dimensions~\cite{ekn08}, then consider the Morita equivalence based on
this constuction.
In Ref.~\cite{ekn08}, the two cases were considered, $\mathcal{N}=(1,1)$ and $\mathcal{N}=(2,2)$.
In the former case, the deformations consistent with supersymmetry turned out to be only
bosonic ones.
In the latter case,
only the so-called $Q$-deformations are consistent with the super
Heisenberg group structure of noncommutative supertori.
Thus here we only consider $\mathcal{N}=(2,2)$ with $Q$-deformation.
The Q-deformation is defined by the Moyal product $*$ with the supercharge $Q_\pm$:
\begin{equation}
\ast
=
\exp\left[\frac{i}{2}\varTheta\epsilon^{\mu\nu}
\overleftarrow{\frac{\partial}{\partial X^{\mu}}}
\overrightarrow{\frac{\partial}{\partial X^{\nu}}}
-\frac{C}{2}\left(\overleftarrow{Q_{+}}\overrightarrow{Q_{-}}
+\overleftarrow{Q_{-}}\overrightarrow{Q_{+}}\right)
\right].
\end{equation}
Here the supercharges $Q_{\pm}$, $\bar{Q}_{\pm}$ are defined by
\begin{align}
Q_{+}&=\frac{\partial}{\partial \theta^{+}}-\frac{\bar{\theta}^{+}}{2}
\left(\frac{\partial}{\partial X^{1}}-i\frac{\partial}{\partial X^{2}}\right)
&
Q_{-}&=\frac{\partial}{\partial \theta^{-}}-\frac{\bar{\theta}^{-}}{2}
\left(\frac{\partial}{\partial X^{1}}+i\frac{\partial}{\partial X^{2}}\right)
\notag\\
\bar{Q}_{+}&=\frac{\partial}{\partial \bar{\theta}^{+}}-\frac{\theta^{+}}{2}
\left(\frac{\partial}{\partial X^{1}}-i\frac{\partial}{\partial X^{2}}\right)
&
\bar{Q}_{-}&=\frac{\partial}{\partial \bar{\theta}^{-}}-\frac{\theta^{-}}{2}
\left(\frac{\partial}{\partial X^{1}}+i\frac{\partial}{\partial X^{2}}\right).
\end{align}
In the operator formalism,
the nontrivial commutation relations
among supercoordinates are given by
\begin{align}
[X^{1},X^{2}]&=i\varTheta-\frac{i}{2}C\bar{\theta^{+}}\bar{\theta^{-}},
&
[X^{1},\theta^{+}]&=\frac{1}{2}C\bar{\theta^{-}},
&
[X^{1},\theta^{-}]&=\frac{1}{2}C\bar{\theta^{+}},
\notag\\
[X^{2},\theta^{+}]&=\frac{i}{2}C\bar{\theta^{-}},
&
[X^{2},\theta^{-}]&=-\frac{i}{2}C\bar{\theta^{+}},
&
[\theta^{+},\theta^{-}]&=C.
\label{n2alg}
\end{align}
We note that the fermionic coordinates $\bar{\theta}^{+},\bar{\theta}^{-}$
appear as the central elements.

 For supertori, there are two types of spin structures, even and odd.
However, for the even spin structure there is no effect from supersymmetrization \cite{ekn08}.
Thus we consider the $\mathcal{N}=(2,2)$ case with odd spin structure.
Setting the noncommutative parameters
 $\varTheta = \frac{1}{2\pi} $ and  $C=1$~\cite{footnote1},
 the generators $U=U_{1}$ and $V=U_{2}$ satisfy
\begin{align}
UX^{\mu}U^{-1}&=X^{\mu}+e_{U}^{\mu},&
U\theta^{\pm} U^{-1}&=\theta^{\pm},&
U\bar{\theta}^{\pm} U^{-1}&=\bar{\theta}^{\pm},
\notag\\
VX^{\mu}V^{-1}&=X^{\mu}+e_{V}^{\mu},&
V\theta^{\pm} V^{-1}&=\theta^{\pm}+\delta^{\pm},&
V\bar{\theta}^{\pm} V^{-1}&=\bar{\theta}^{\pm}+\bar{\delta}^{\pm},
\label{translation}
\end{align}
where the supercoordinates obey the algebra \eqref{n2alg}
and the lattice vectors $e_{U}^{\mu}$ and $e_{V}^{\mu}$ are given by
\begin{align}
e_{U}^{\mu}&={}^{t}(1,0),\notag\\
e_{V}^{\mu}&={}^{t}
\bigl(\Re(\tau+\bar{\theta}^{+}\delta^{+}+\theta^{+}\bar{\delta}^{+}),\,
\Im(\tau+\bar{\theta}^{+}\delta^{+}+\theta^{+}\bar{\delta}^{+})\bigr).
\end{align}
Then the explicit form of $U,\,V$ can be obtained as
\begin{align}
U&=\exp(2\pi is),
\notag\\
V&=\exp\biggl[2\pi i(\Re\,\tau)s+(\Im\,\tau)\frac{\partial}{\partial s}
+\delta^{+}\eta+\delta^{-}\frac{\partial}{\partial\eta}\biggr].
\label{n2genq}
\end{align}
The commutation relation between $U$ and $V$ is
given by
\begin{equation}
UV=\exp(-2\pi i\,\Im\,\tau)VU.
\end{equation}
The embedding map 
can be written from
\eqref{n2genq} as
\begin{equation}
\widetilde{\Phi}_{Q}=\,\bordermatrix{&U&V\cr
\frac{\partial}{\partial s}&0&\Im\,\tau\cr
s&1&\Re\,\tau\cr
\frac{\partial}{\partial \eta}&0&\delta^{-}\cr
\eta&0&\delta^{+}\cr
}.
\end{equation}
We can compare this with the bosonic case where the noncommutativity parameter
$\theta_{ij}$ is given by \eqref{cyctheta}. If we write $U:= U_{\vec{E}_1}$ and $V:=U_{\vec{E}_2}$
with supersymmetric basis vectors $\vec{E}_1$ and $ \vec{E}_2$,
then the relation \eqref{cocycler}  becomes
\begin{eqnarray}
 U_{\vec{E}_1} U_{\vec{E}_2}=   e^{2\pi i \widetilde{\theta}_{12}}U_{\vec{E}_2} U_{\vec{E}_1} ,
\label{susy_cocycle}
\end{eqnarray}
and the relation
 \eqref{cyctheta}
becomes
 \begin{align}
\widetilde{\theta}_{12} = \vec{E}_1 \cdot  \widetilde{J}_0  \vec{E}_2  \ \ {\rm where } \  \ \widetilde{J}_0 =\begin{pmatrix}
\begin{pmatrix} 0 & I \\ -I & 0 \end{pmatrix} & 0 \\ 0 &
\begin{pmatrix} 0 & I \\ I & 0 \end{pmatrix}
\end{pmatrix}.
\label{susy_cyctheta}
\end{align}
We relegate the higher dimensional extension of this construction to the appendix.



We now consider Morita equivalence of noncommutative
supertori in the ${\R}^p$ type embedding case.
We consider the Grassmann algebra ${\cal V}$ over ${\C}$,
and decompose ${\cal V}$ into even  and odd parts,
${\cal V}={\cal V}^0 \bigoplus {\cal V}^1$.
The noncommutativity parameters $\widetilde{\Theta}$ for noncommutative supertori
 belongs to ${\cal V}^0$.
Rather than follow the approach used in the previous section,
here we search for symmetry which yields  Morita equivalence.
We know from the previous section that the endomorphism algebra of the module of noncommutative torus
 is Morita equivalent to the given noncommutative torus and the condition is given by \eqref{D-pub}.
This can be translated into
\begin{equation}
 \Phi^t J_0 \Phi' = K ,
\label{b_dual}
\end{equation}
where $\Phi$ is the embedding map of the given torus and $\Phi'$ is the embedding map of
the dual torus, and $K$ is an $n \times n$ matrix whose elements belong to ${\Z}$.
The above relation
can be put into the following form in the supersymmetric case:
\begin{equation}
\widetilde{\Phi}^t \widetilde{J}_0 \widetilde{\Phi}' = B^t J_0  B' + F^t \hat{J}_0 F' = \widetilde{K}  \ \ {\rm where} \ \
J_0=\begin{pmatrix} 0 & I_p \\ -I_p & 0 \end{pmatrix}, \
\hat{J}_0 =
\begin{pmatrix} 0 & I_r \\ I_r & 0 \end{pmatrix},
\label{s_dual}
\end{equation}
where
 $\widetilde{\Phi}:=\begin{pmatrix} B \\ F \end{pmatrix}$ and
 $\widetilde{\Phi}':=\begin{pmatrix} B' \\ F' \end{pmatrix}$ are
 the embedding maps (here $B,  F  $ represent bosonic and fermionic parts)
  of the given supertorus and the dual supertorus, respectively,
 and the elements of the matrix $\widetilde{K}$ belong to $ {\Z}$.
Note that the entries of  $B, B'$ and $F, F'$ belong to ${\cal V}^0$ and ${\cal V}^1$, respectively.
Below we denote the antisymmetric noncommutativity parameter matrix $\widetilde{\Theta}$
and show that it contains both bosonic and fermionic contributions
even though $\widetilde{\Theta}$ itself belongs to ${\cal V}^0$:
\begin{equation}
\widetilde{\Theta}= \widetilde{\Phi}^t \widetilde{J}_0 \widetilde{\Phi} = B^t J_0 B + F^t \hat{J}_0 F.
\label{theta_BF}
\end{equation}
Note that under the change of basis,
the matrix $\widetilde{K}$ in the duality condition \eqref{s_dual}
 can be any element in $GL(n,{\Z})$.

 From the condition  \eqref{s_dual} we can express the bosonic part of the dual embedding map as
\begin{equation}
B'= - J_0 B^{-t}(\widetilde{K}-F^t \hat{J}_0 F').
\label{dual_b}
\end{equation}
 Thus using the relation  \eqref{theta_BF}
we can express the noncommutativity matrix $\widetilde{\Theta}' $ of the dual supertorus as
\begin{eqnarray}
\widetilde{\Theta}'  &=& -(\widetilde{K}-F^t \hat{J}_0 F')^t
B^{-1}J_0 J_0 J_0
B^{-t}(\widetilde{K}-F^t \hat{J}_0 F') +  F^{'t} \hat{J}_0 F' ,
\nonumber \\
&=& -(\widetilde{K}-F^t \hat{J}_0 F')^t (B^t J_0 B)^{-1}
(\widetilde{K}-F^t \hat{J}_0 F')
+F^{'t} \hat{J}_0 F',
\label{s_btheta}
\end{eqnarray}
where we used
 $J_0^{-1} = - J_0$.
We can also express $\widetilde{\Theta}' $ directly using the dual embedding map $\widetilde{\Phi}'$.
 From the relation \eqref{s_dual} we have
\begin{eqnarray}
\widetilde{\Phi}' = (\widetilde{\Phi}^t \widetilde{J}_0 )^{-1} \widetilde{K},
\label{dual_map}
\end{eqnarray}
thus
\begin{eqnarray}
\widetilde{\Theta}' & =& {\widetilde{\Phi}}^{'t} \widetilde{J}_0 {\widetilde{\Phi}}^{'} \nonumber \\
  & = & \widetilde{K}^t (\widetilde{\Phi}^t \widetilde{J}_0 )^{-t} \widetilde{J}_0
  (\widetilde{\Phi}^t \widetilde{J}_0 )^{-1} \widetilde{K} \nonumber \\
  & = & \widetilde{K}^t (\widetilde{\Phi}^t \widetilde{J}_0^t
  \widetilde{\Phi})^{-1} \widetilde{K}.
\label{dual_stheta}
\end{eqnarray}
Since
\[
\widetilde{J}_0^t = \begin{pmatrix} J_0 & 0\\
0 & \hat{J}_0 \end{pmatrix}^t = \begin{pmatrix} -J_0 & 0\\
0 & \hat{J}_0 \end{pmatrix},
\]
we can write
$\widetilde{\Phi}^t \widetilde{J}_0^t   \widetilde{\Phi} = - \Theta_B + \Theta_F$
where we denote $\Theta_B := B^t J_0 B $  and $ \Theta_F := F^t \hat{J}_0 F $.
Therefore
\eqref{dual_stheta} can be written as
\begin{eqnarray}
\widetilde{\Theta}'
  & = & \widetilde{K}^t (- \Theta_B + \Theta_F )^{-1} \widetilde{K} \nonumber \\
  & = & - \widetilde{K}^t \widehat{\Theta}_b^{-1} \sum_{m=0}^\infty
  (\widehat{\Theta}_s \widehat{\Theta}_b^{-1})^m  \widetilde{K}
  ,
\label{inv_dstheta}
\end{eqnarray}
where $\widehat{\Theta}_b$ is the body part of $\Theta_B$ and $\widehat{\Theta}_s$ is
the soul part of $ \Theta_F - \Theta_B$.
Note that the body parts of $\Theta_B$ and $\widetilde{\Theta}$ are the same.


Now, let us consider
$\sigma_n=\begin{pmatrix} 0 & I_n  \\ I_n &
0 \end{pmatrix}$ where $n$ is the dimension of the torus.
Its action is  given by  Eq. \eqref{gaction_theta},
\begin{equation}
 \sigma_n \widetilde{\Theta} =  \widetilde{\Theta}^{-1} = ( \Theta_B + \Theta_F )^{-1}
 = \widetilde{\Theta}_b^{-1} \sum_{m=0}^\infty (- \widetilde{\Theta}_s \widetilde{\Theta}_b^{-1})^m ,
\label{s_sigman}
\end{equation}
where $\widetilde{\Theta}_b, \; \widetilde{\Theta}_s$ are the body and soul parts of
 $\widetilde{\Theta}$, respectively.
 Since $\widehat{\Theta}_b^{-1}$ in \eqref{inv_dstheta} and $\widetilde{\Theta}_b^{-1}$ are the same,
 $\widetilde{\Theta}^{-1}$ is just differ from $\widetilde{\Theta}'$
in the soul part up to
the action of  $\widetilde{K} \in GL(n,{\Z})$
which generates Morita equivalent torus as we see below.
The overall minus sign can be absorbed by the sign flip of a half of the
basis vectors as in the bosonic case.

 The relation \eqref{s_btheta} tells us that the soul part $F'$ of the dual map
  is not restricted by the duality condition \eqref{s_dual}.
Namely, if  the two $\widetilde{\Theta}$'s have the same body parts and differed by the soul parts belonging
 to ${\cal V}^0$,
then the two corresponding tori are Morita equivalent.
Thus we can say that $\sigma_n$ generates Morita equivalent torus.

The above statement that the same body parts up to elements in
${\cal V}^0$ yield
equivalent tori dictates us  another  symmetry action of the following element
 of $SO(n,n,{\cal V}_{\Z}^0)$ \cite{footnote2},
\[
\nu(\widetilde{N})=\begin{pmatrix}
I_n & \widetilde{N}\\
0 &I_n\\
\end{pmatrix}
\]
where $\widetilde{N}$
is an antisymmetric  $n \times n$ matrix whose entries are in ${\cal V}_{\Z}^0$
and ${\cal V}_{\Z}^0$ denotes Grassmann even number
whose body part belongs to ${\Z}$.
 The action of $\nu(\widetilde{N})$ is given as before by \eqref{gaction_theta}
\begin{equation}
 \nu(\widetilde{N}) \widetilde{\Theta} = \widetilde{\Theta} + \widetilde{N} .
\end{equation}

The ``rotation" by $\rho(\tilde{R}) \in SO(n,n,{\cal V}_{\Z}^0)$ can be similarly considered as in the bosonic case.
We consider $\rho(\tilde{R}) \in SO(n,n,{\cal V}_{\Z}^0)$ given by
\[ \rho(\tilde{R})=\begin{pmatrix}
\tilde{R}^t & 0\\
0 & \tilde{R}^{-1}\\
\end{pmatrix},
\]
where $\tilde{R} \in GL(n, {\cal V}_{\Z}^0)$.
%
%
When a  basis $\{\vec{E}_i \}$ ($i=1,2, \cdots, n$) is given, we can consider
a general embedding vector $\vec{X}$ in terms of given basis such as
$\vec{X}= \sum_1^n X_i \vec{E}_i$ where
 $X_i \in {\cal V}_{\Z}^0 $, and $\vec{E}_i$'s satisfy the relation \eqref{s_theta}.
%
%
Then $\widetilde{\theta}_{XY}$ for two general embedding vectors $\vec{X}$ and $\vec{Y}$
is given by
  \begin{equation}
 \label{gs_theta}
\widetilde{\theta}_{XY} = \sum_{i,j=1}^n ( X_i \vec{E}_i) \cdot   \widetilde{J}_0  \ ( Y_j \vec{E}_j )
  = \sum_{i,j=1}^n X_i \widetilde{\theta}_{ij} Y_j .
 \end{equation}
Namely,
 \begin{equation}
 \label{gcs_theta}
U_X U_Y = \exp( 2 \pi i \, X^t  \widetilde{\Theta} Y) U_Y U_X ,
 \end{equation}
where  $X=(X_1, \dots, X_n)$ and $ Y=(Y_1,\dots, Y_n)$.
Under the action of $\rho (\tilde{R})$, the transformed $X$ is given by $ X'=  \rho(\tilde{R}) X = \tilde{R}^t X$,
and thus
  \begin{equation}
 \label{trstheta}
U_{X'} U_{Y'}  = \exp( 2 \pi i \, (\tilde{R}^t X)^t \widetilde{\Theta} (\tilde{R}^t Y)) U_{Y'} U_{X'}
  =  \exp( 2 \pi i \, X^t \tilde{R} \widetilde{\Theta} \tilde{R}^t Y)U_{Y'} U_{X'} .
 \end{equation}
Therefore, the two tori with $\widetilde{\Theta}$ and $\widetilde{\Theta}' = \tilde{R} \widetilde{\Theta} \tilde{R}^t $ are isomorphic.
 In \eqref{s_btheta} we can see that the action of $\rho (\tilde{R})$ is already incorporated in the transformation to the dual
 torus.  There we see that $\tilde{R}$ appears as $(\widetilde{K}-F^t \hat{J}_0 F')^t$ acting on the body part of
 $ \sigma_n \widetilde{\Theta}$. Since $\widetilde{\Theta}$'s with the same body part are Morita equivalent,  $\rho (\tilde{R}) {\Theta_B}^{-1}$
 in  \eqref{s_btheta} is equivalent to $\rho (\tilde{R}) \widetilde{\Theta}^{-1}$.

So far, we have shown that the three elements
$\rho (\tilde{R}), \nu (\widetilde{N})$ with $\tilde{R}, \widetilde{N} \in {\cal V}_{\Z}^0$, and $\sigma_{n}$
 yield Morita equivalent noncommutative $n$-supertori.
 When $n=2$,
  the group $SO(2,2, {\cal V}_{\Z}^0)$ is generated by these elements,
  $\rho (\tilde{R}), \nu (\widetilde{N})$ and $\sigma_{2}$.
Therefore, we have the following result in the two dimensional case.


{\bf Theorem}:   If $g \in SO(2,2,{\cal V}_{\Z}^0)$,
then the noncommutative
supertorus corresponding to $g \widetilde{\Theta}$ is Morita equivalent to the
noncommutative supertorus corresponding to $\widetilde{\Theta}$ in two dimensions.
\\

\section{Conclusion}

In this paper, we show that the group $SO(2,2,{\cal V}_{\Z}^0)$ yields Morita equivalent
noncommutative supertori in the two dimensional case.
 For the higher dimensional case ($n>2$),
  we obtain the three elements belonging to the group $SO(n,n,{\cal V}_{\Z}^0)$, which are
$ \rho (\tilde{R}), \nu(\widetilde{N}) $ and $\sigma_n$ that yield Morita equivalent tori.
However, due to the absence of  ${\R}^p \times {\Z}^q$ type embedding construction with nonzero $q$
in the supersymmetric case we do not have $\sigma_k$ with $k<n$.
  Thus  for the moment, we are short of having $SO(n,n,{\cal V}_{\Z}^0)$ symmetry group
 in the higer dimensional case.
\\

%
\noindent
{\Large \bf Acknowledgments}

\vspace{5mm} \noindent The authors thank KIAS for hospitality
during the time that this work was done.
This work was supported
by the National Research Foundation(NRF) of Korea grants
funded by the Korea government(MEST), 2009-0075129(E.C.-Y.)
and 2009-0070957(H.K.),
and is a result of research
activities (Astrophysical Research Center for the Structure and
Evolution of the Cosmos (ARCSEC)) by grant No.\
R01-2006-000-10965-0 from the Basic Research Program supported by
KOSEF(H.N.).
\\

\section*{Appendix:\\
 Noncommutative supertori in higher dimensions}

Here we consider the construction of noncommutative supertori in higher dimensions with Q-deformation.
 Since the extended supersymmetries can be obtained from the higher dimensional ones
by dimensional reductions, here we only consider the higher dimensional ones.
 First of all, we like to recall that the supersymmetrization of
noncommutative torus with the embedding map of ${\R}^p \times {\Z}^q$ type,
where the dimension of torus is $n=2p+q$ with nonzero $q$, is as yet unknown.
This is because the supersymmetrization of noncommutative torus with nonzero $q$ embedding case
necessarily deals with a deformation of discrete(lattice) supersymmetry, and this is not well
understood  so far \cite{footnote3}.
Thus here we only consider the vanishing $q$ case, namely  the supersymmetrization of
noncommutative tori with ${\R}^p$ type embedding map.

In order to respect the property of Heisenberg group, we consider noncommutative
supertori in higher dimensions with only odd spin structures and
with Q-deformations.
In this case, we can express the basis vectors in an embedding map
as $\vec{E}_i =(x, \alpha) \in {\R}^{n|m}$ where $i =1, \dots, n$ and $x, \ \alpha$ are
bosonic(Grassmann even) and fermionic(Grassmann odd) variables respectively.
Notice also that in order to satisfy the structure of Heisenberg group we only consider the cases
in which both $n$ and $m$ are even integers, say $n=2p$ and $m=2r$.
Now the noncommutativity parameter can be expressed as
\begin{equation}
\widetilde{\theta}_{ij} =   \vec{E}_i \cdot  \widetilde{J}_0  \vec{E}_j  \ \ {\rm where } \  \ \widetilde{J}_0 =\begin{pmatrix}
\begin{pmatrix} 0 & I_p \\ -I_p & 0 \end{pmatrix} & 0 \\ 0 &
\begin{pmatrix} 0 & I_r \\ I_r & 0 \end{pmatrix}
\end{pmatrix}.
\label{s_theta}
\end{equation}
Therefore the antisymmetric noncommutativity parameter matrix $\widetilde{\Theta}$ is given by
\begin{equation}
\widetilde{\Theta} = \widetilde{\Phi}^t \widetilde{J}_0   \widetilde{\Phi},
\label{susyTheta}
\end{equation}
as in the bosonic case \eqref{Theta}.
Note that the embedding map $\widetilde{\Phi}$ can be expressed in terms of basis vectors
and can be decomposed of bosonic and fermionic parts as
\begin{equation}
\widetilde{\Phi} = \begin{pmatrix}
\vec{E}_1  &\vec{E}_2  & \cdots  & \vec{E}_n
\end{pmatrix} :=  \begin{pmatrix}
B \\ F \end{pmatrix},
\label{sembed}
\end{equation}
where $B$ is the bosonic part of the embedding map given by  a $2p \times 2p$  matrix (here $n=2p$) and $F$ is
the fermionic part of the embedding map given by a $2r \times 2p$ matrix.
Then \eqref{susyTheta} can be decomposed as
\begin{equation}
\widetilde{\Theta} = B^t J_0  B + F^t \hat{J}_0 F  \ \ {\rm where} \ \
J_0=\begin{pmatrix} 0 & I_p \\ -I_p & 0 \end{pmatrix}, \
\hat{J}_0 =
\begin{pmatrix} 0 & I_r \\ I_r & 0 \end{pmatrix} .
\label{dec_susyTheta}
\end{equation}

In principle,  in the ${\R}^p$ type embedding case there seems to be no obstruction to construct
noncommutative supertori in higher dimensions. Namely,  using the operators
$s$, $\frac{\partial}{\partial s}$, etc. and (a part of) supercharges
which satisfy super Heisenberg algebra, we can construct
noncommutative supertori.
In general, we should consider some additional coordinates on superspace such as
harmonic superspace~\cite{GIOS} or projective superspace~\cite{Karlhede:1984vr}, if we
want to contain  supersymmetric field theory  in higher dimensions.
We leave this extended construction as an open issue.
\\





\begin{thebibliography}{88}
\bibitem{footnote1} This normalization is different
 from our previous  work~\cite{ekn08} by a factor of $2\pi$.
\bibitem{footnote2} For the soul part of $\tilde{N}$, this can be directly shown
by using \eqref{s_btheta} twice and arbitrariness of $F'$.
\end{thebibliography}
\end{document}